\documentclass[preprint,prl,superscriptaddress,showpacs,amsmath,amssymb]{revtex4}
\usepackage{graphicx}
\usepackage{dcolumn}
\usepackage{bm}

\begin{document}


\title{Observation of an energetic radiation burst from mountain-top thunderclouds}

\author{H. Tsuchiya}
\affiliation{Cosmic Radiation Laboratory, Riken, 2-1, Hirosawa, Wako, 
Saitama 351-0198, Japan}
\author{T. Enoto}
\affiliation{Department of Physics, University of Tokyo, 7-3-1, Hongo, 
Bunkyo-ku, Tokyo 113-0033, Japan}
\author{T. Torii}
\affiliation{Tsuruga Head Office, Japan Atomic Energy Agency, 2-1, Shiraki, 
Tsuruga, Fukui 919-1279, Japan}
\author{K. Nakazawa}
\affiliation{Department of Physics, University of Tokyo, 7-3-1, Hongo, 
Bunkyo-ku, Tokyo 113-0033, Japan}
\author{T. Yuasa}
\affiliation{Department of Physics, University of Tokyo, 7-3-1, Hongo, 
Bunkyo-ku, Tokyo 113-0033, Japan}
\author{S. Torii}
\affiliation{Department of Physics, University of Tokyo, 7-3-1, Hongo, 
Bunkyo-ku, Tokyo 113-0033, Japan}
\author{T. Fukuyama}
\affiliation{Department of Physics, University of Tokyo, 7-3-1, Hongo, 
Bunkyo-ku, Tokyo 113-0033, Japan}
\author{T. Yamaguchi}
\affiliation{Department of Physics, University of Tokyo, 7-3-1, Hongo, 
Bunkyo-ku, Tokyo 113-0033, Japan}
\author{H. Kato}
\author{M. Okano}
\affiliation{Cosmic Radiation Laboratory, Riken, 2-1, Hirosawa, Wako, 
Saitama 351-0198, Japan}
\author{M. Takita}
\affiliation{Institute for Cosmic  Ray Research, University of Tokyo, 
Kashiwa, Chiba, 277-8582, Japan}
\author{K. Makishima}
\affiliation{Cosmic Radiation Laboratory, Riken, 2-1, Hirosawa, Wako, 
Saitama 351-0198, Japan}
\affiliation{Department of Physics, University of Tokyo, 7-3-1, Hongo, Bunkyo-ku, 
Tokyo 113-0033, Japan}

\date{\today}
\begin{abstract}
During thunderstorms on 2008 September 20, a simultaneous detection of $\gamma$ rays 
and electrons was made at a mountain observatory in Japan located 2770 m above sea level.
Both emissions, lasting 90 seconds, were associated with thunderclouds rather than lightning.
The photon spectrum,  extending to 10 MeV, can be interpreted 
as consisting of bremsstrahlung $\gamma$ rays arriving from a source which is
$60 - 130$ m in distance at 90\% confidence level. The observed electrons
are likely to be dominated by a primary population escaping from an acceleration region 
in the clouds.
\end{abstract}

\maketitle

\section{Introduction\label{sec:intro}}
Energetic radiation bursts from thunderstorm activity have been observed by space 
observatories as well as  ground-based detectors. Interestingly, these bursts appear to be classified
into two types according to their duration. One comprises short bursts lasting for milli-seconds or less,
mainly observed  from the upper atmosphere of the 
Earth~\cite{BATSE,RHESSI}, natural lightning discharges~\cite{moore,Dwyer_natural_2005},
and rocket-triggered ones~\cite{Dwyer_exp_science,Dwyer_exp_typspe,Dwyer_exp_10MeV}.  
Though not necessary homogeneous, 
they generally occur in association with lightning discharges.
The other consists of long bursts with duration of a few seconds to a few minutes,
mostly detected at aircraft altitudes~\cite{MP_1985,Eack_1996}, at high mountains~\cite{SUSZ_1996,EAS_2000,Chub_2000,Alex_2002,nr_muraki_2004}, 
and around the coastal area of Japan Sea~\cite{monju_2002,growth_2007}.
Unlike the short bursts, few of them appear to be clearly
associated with lightning discharges.  

To date, at least some of these radiation bursts are thought to result from relativistic electrons, 
which are produced in turn by a mechanism involving runaway and avalanches of electrons in the atmosphere.  
The mechanism, first proposed by \citet{RREA_model_1992}, needs intense electric 
fields, high-energy seed electrons, and spatial length long enough for the avalanches to develop. 
When these conditions are fulfilled, the electrons can gain energy from the electric 
fields fast enough to exceed their ionization loss. Then,  they will be accelerated to 
relativistic energies, and produce bremsstrahlung $\gamma$ rays.  

Actually, $\gamma$-ray spectra extending to $\sim$10 MeV have been observed from 
both short and long bursts~\cite{Dwyer_exp_10MeV,RHESSI,growth_2007},
with their properties generally consistent with the bremsstrahlung emission scenario.
However,  according to Monte Carlo simulations considering a feedback 
mechanism in which back-scattered photons and/or positrons successively supply seed 
electrons in the high-electric-field region, electron avalanches are likely to last only for
microseconds~\cite{Dwyer_MC2003, Babich_MC, Dwyer_MC2007}.
Observationally, only a few examples of long bursts have so far been reported
to extend to MeV energies. Therefore, it is not yet clear whether the runaway 
electron avalanche scenario, which is generally successful on some short 
bursts~\cite{RHESSI,Dwyer_exp_10MeV},
can also explain the prolonged events.
Furthermore, we need to explain the fact that some long bursts consist solely of x/$\gamma$ rays~\cite{MP_1985,Eack_1996,SUSZ_1996,EAS_2000,Chub_2000,monju_2002,growth_2007}, 
while others involve only charged particles~\cite{Alex_2002,nr_muraki_2004}. 
Here, utilizing a detector that is capable of separating photons from charged particles, 
we first report on a simultaneous detection of $\gamma$ rays and electrons in one long burst, 
with the photon spectrum extending to 10 MeV.
\section{Experiment\label{sec:exp}}
The Norikura cosmic-ray observatory ($36^\circ6'$N, $137^\circ33'$E), belonging to 
the Institute for Cosmic Ray Research, the University of Tokyo, is located on a 
high mountain in Gifu prefecture, Japan. Due to its high altitude, 2770 m above 
sea level, 
as well as meteorological  conditions from summer to autumn,  the observatory 
is frequently shrouded in thunderclouds. 

Outside the observatory, we installed a radiation detection system 
on 2008 September 4, and operated it until 2008 October 2. 
The system consists of a spherical NaI scintillator with a diameter of 7.62 cm, 
and a $45\times40\times0.5\, \mathrm{cm^3}$ plastic scintillator which 
is enclosed in an aluminum box with the top and 
bottom being 1 mm and 3 mm thick, respectively.
Each scintillator has a photomultiplier (HAMAMATSU R878) of its own,  and each output 
signal is fed to a self-triggering electronics system, incorporating a 12 bit 
analog-to-digital converter (ADC). 
The detected events are collected in the order of arrival without individual arrival-time information, 
and recorded every 1 sec synchronized by Global Positioning System.
Thus, both scintillators have a time resolution of 1 sec.  
The NaI scintillator was operated over $10$ keV $-12$ MeV, while the plastic scintillator,
placed above it, was operated with a threshold energy deposit of 500 keV.  

The plastic scintillator has a high sensitivity for charged particles while it is much 
less sensitive to photons. Therefore, when the NaI detector is triggered by 
an incident particle without a coincident (within 0.7 $\mu$sec) signal in the plastic scintillator, 
the incident particle would be a photon, not a charged particle. 
Thus, the plastic signals mainly represent incoming charged particles including electrons, 
and using them in anti-coincidence, the NaI signals can be considered to
indicate $\gamma$ rays with little ($\sim$ 0.1\%) contamination by electrons.

Aiming at  an independent radiation observation,
we installed, inside the observatory, another inorganic scintillation 
detector. 
This indoor detector uses a cubic bismuth germanate (BGO) scintillator of 
5 $\times$ 5 $\times$ 15 $\mathrm{cm^3}$, which is coupled to a photomultiplier (HAMAMATSU R7600U-203), 
and observes in the 100 keV $-$ 5 MeV range. The signal is pulse-height analyzed by a self-triggering 8 bit
ADC, and the pulse height is recorded, on event-by-event basis, with arrival-time information
of 20 ms resolution.

In addition to the radiation observations, we measure optical and electric field variations outside
the observatory. 
The visible-light intensity is measured by an optical sensor
using a photodiode (HAMAMATSU S1226-8BK) having a sensitivity 
to photons with wavelength of 320 $-$ 1000 nm, with its peak at 750 nm.
Output signals of this optical sensor are fed 
to a 12 bit ADC every 1 sec, and 
recorded as voltages between 0 and +10 V with a resolution of 5 mV. 
The electric field is measured by a commercial electric field mill (BOLTEK EFM-100). 
Its outputs are also collected by a 12 bit ADC every 1 sec,
and converted to the electric field strength between $\pm$ 100 $\mathrm{kVm^{-1}}$ 
with a resolution of 50 $\mathrm{Vm^{-1}}$.

\section{Results}\label{sec:res}
\begin{figure}
\includegraphics[width=1.0\textwidth]{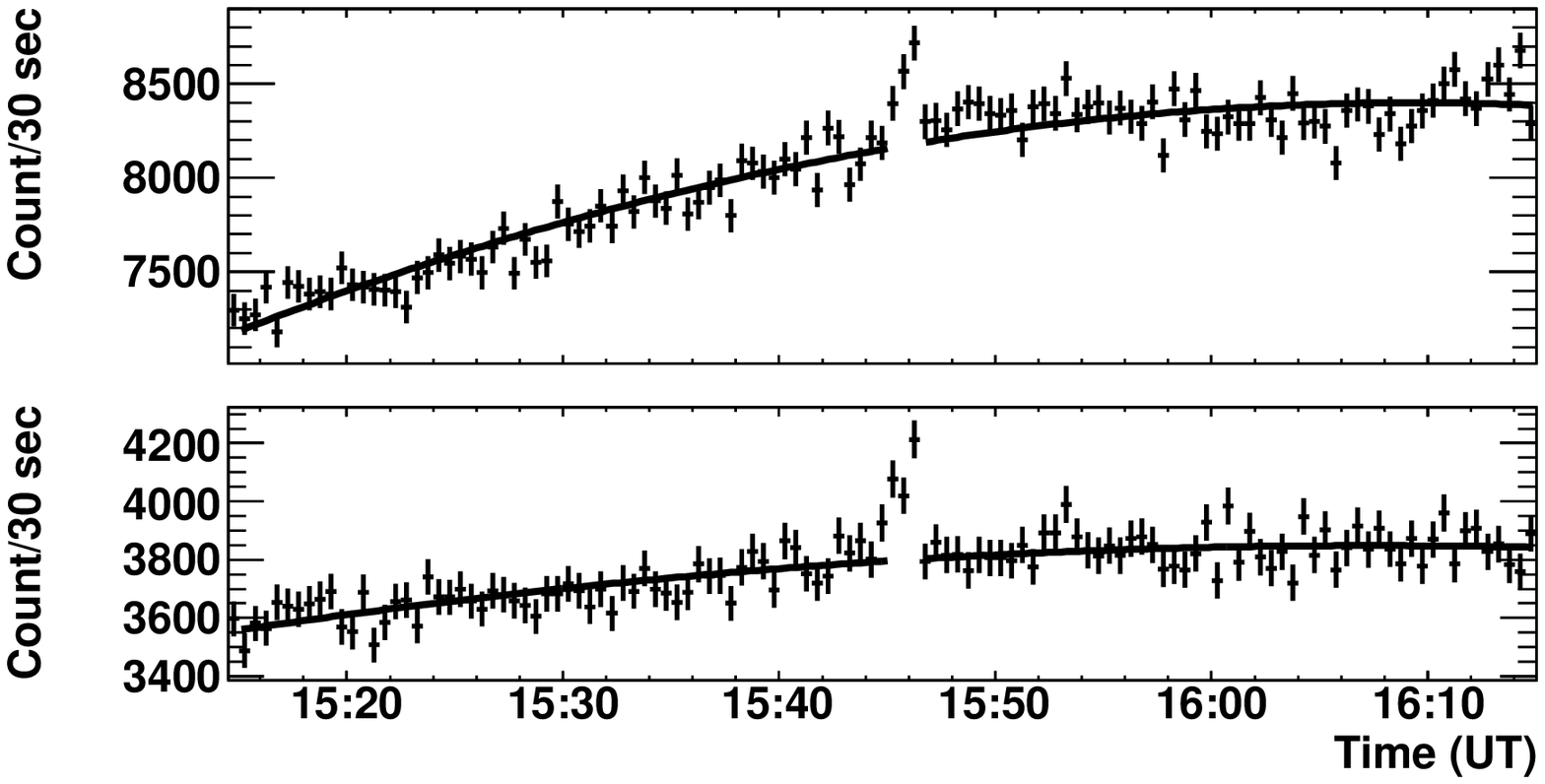}
\caption{%
Count rates per 30 seconds of $>$ 10 keV NaI (top) and $>$ 500 keV 
plastic (bottom) scintillators between 15:15 and 16:15 UT on 2008 September 20. 
The horizontal axis represents universal time, and all errors are statistical 1$\sigma$. 
Each solid curve shows estimated background level.}
\label{fig:nai_pl}
\end{figure}

Examining the data over a period of September 4 to October 2, we
found a long-duration enhancement during thunderstorms on September 20,
but found no other events of similar properties.
Figure~\ref{fig:nai_pl} shows count histories of $>$ 10 keV NaI and $>$ 500 keV plastic scintillators, between 
15:15 and 16:15 UT on September 20. Besides gradual count-rate changes 
mainly due to Radon rain-outs, both
scintillators detected a significant count enhancement, or a burst,  over 
a 90 sec interval of 15:45:10 $-$ 15:46:40 UT. 
Excluding the data obtained in this interval and applying a quadratic least square fitting to 
the remaining data,  we estimated background (solid curves in Fig.~\ref{fig:nai_pl}) in the two scintillators.
Subtracting the interpolated background from the total observed counts
in the 90 sec period and calculating the background errors using the derived 
quadratic function and the number of bins used in the fit procedure, 117, 
both NaI and plastic scintillators gave highly significant increases
of $1208 \pm 150\, (8.1\sigma)$ and $658 \pm 104\, (6.3\sigma)$, respectively.

\begin{figure}
\includegraphics[width=1.0\textwidth]{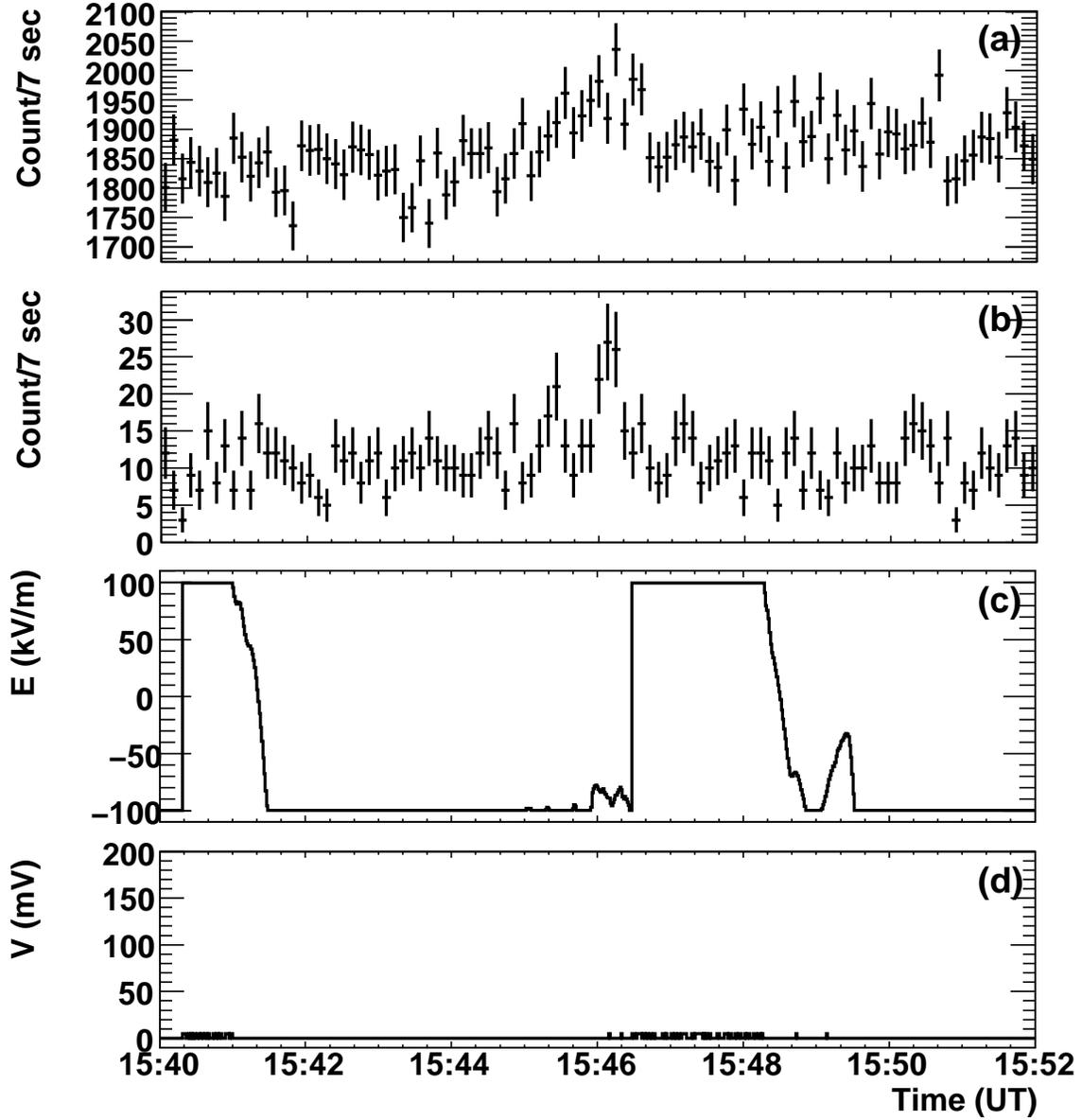}
\caption{%
Detailed count histories of the NaI counter with anticoincidence, and 
the electric field and light sensors, between 15:40 and 15:52 UT. All abscissa are 
universal time. Panels (a) and (b) show count histories per 7 seconds of the NaI counter in $0.01 - 3 $
MeV and $3 - 12$ MeV energies, respectively, with $1\sigma$ statistical errors.
(c) One-second electric field data variations. (d) One-second optical data outputs.}
\label{fig:detail_lc}
\end{figure}

Figure~\ref{fig:detail_lc} gives details of the count histories of the NaI scintillator,
together with outputs of the electric field and light sensors. 
Since the plastic scintillator signals are used here in anticoincidence, 
the NaI events are thought to represent energetic photons.
The excess counts in the $0.01 - 3$ MeV and  $3 - 12$ MeV bands [Fig.~\ref{fig:detail_lc}(a) and (b)] 
become $1033 \pm 159$ and  $91 \pm 15$, respectively, while 
those without anticoincidece in the two energy 
bands are $1030 \pm 159$ and $92 \pm 16$, respectively.
Since the excess counts with and without the anticoincidence thus statistically agree, 
we infer that the NaI signals were dominated by photons, not charged particles.

On this occasion,  the indoor
BGO detector also showed a moderate increase in $>$ 2.6 MeV energies,
with a $\sim 4 \sigma$ significance when summed over a 15 sec interval. 
The absolute time of this increase is consistent  with that in the outdoor
detector, within a relatively large uncertainty of 3 min with which the 
clock of the data-taking computer of the indoor detector was adjusted. 
Supposing that the outdoor and indoor detectors 
observed the same event, the lower significance in the latter, compared to the former, 
suggests that the burst source was located outside the building, and the 
radiation was attenuated by the roofs and/or walls. 

Due to a limited dynamic range of the electric field mill, the electric field [Fig.~\ref{fig:detail_lc}(c)] was 
saturated at $ - 100$ $\mathrm{kVm^{-1}}$ 
at 15:45:10 UT (burst onset), and kept negative polarity until 15:46:29 UT.
At that time, the electric field rapidly changed its polarity from negative to 
positive within 1 sec, and saturated at $+ 100$ $\mathrm{kVm^{-1}}$.  
However, the outputs of the optical sensor [Fig.~\ref{fig:detail_lc}(d)]
remained at 0 mV which is typical at midnight (15:00 UT corresponds to local midnight): 
its small fluctuations, $\sim 5\,\, \mathrm{mV}$, 
reflect the ADC resolution. In reality, when the optical sensor detects lightning,
it typically records a few hundred mV or even $>$ 5 V on rare occasions.  
Thus, we conclude that no lightning occurred during the 
burst, and hence this burst is associated with thunderclouds.

\begin{figure}
\includegraphics[width=0.45\textwidth]{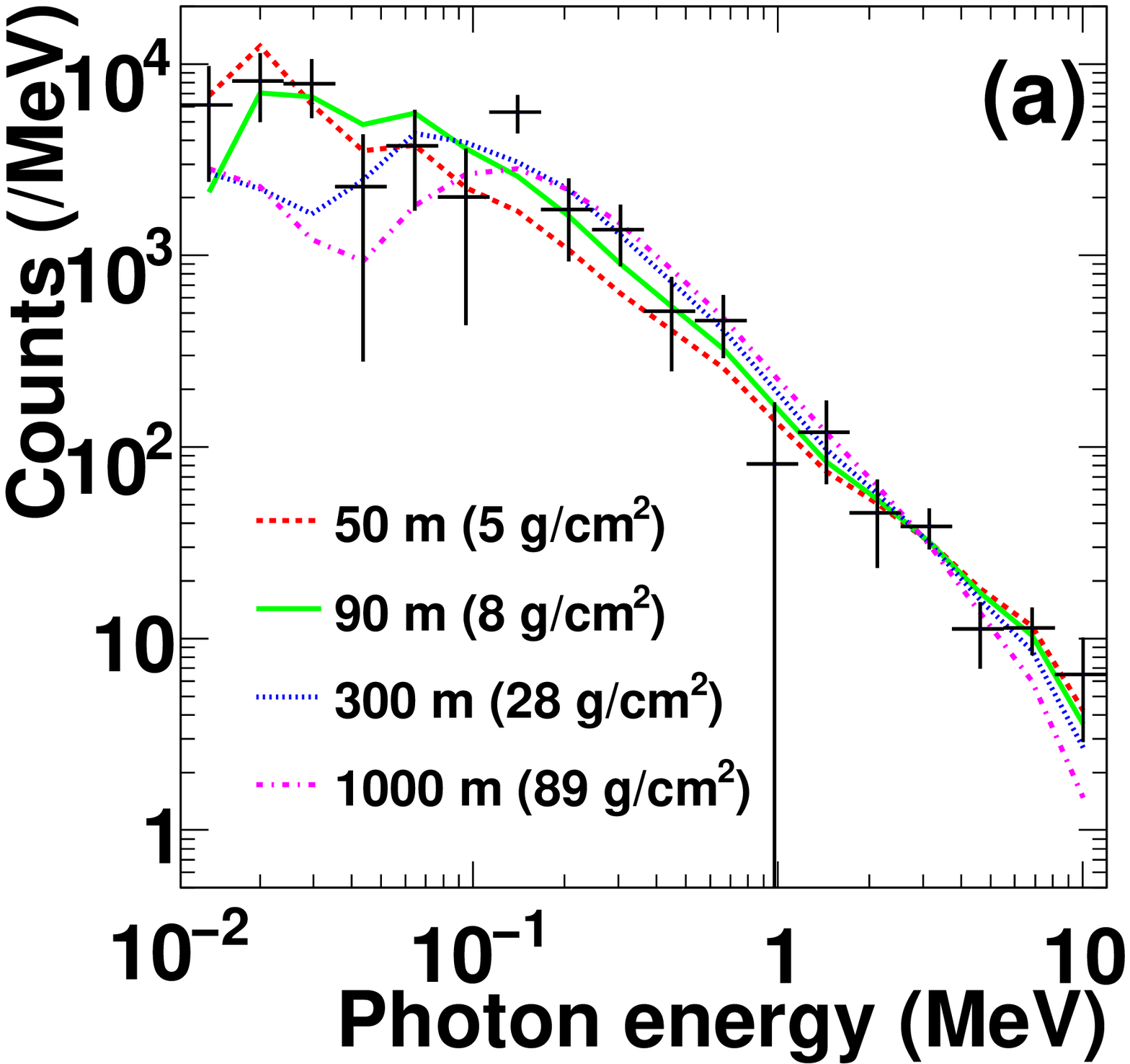}
\includegraphics[width=0.45\textwidth]{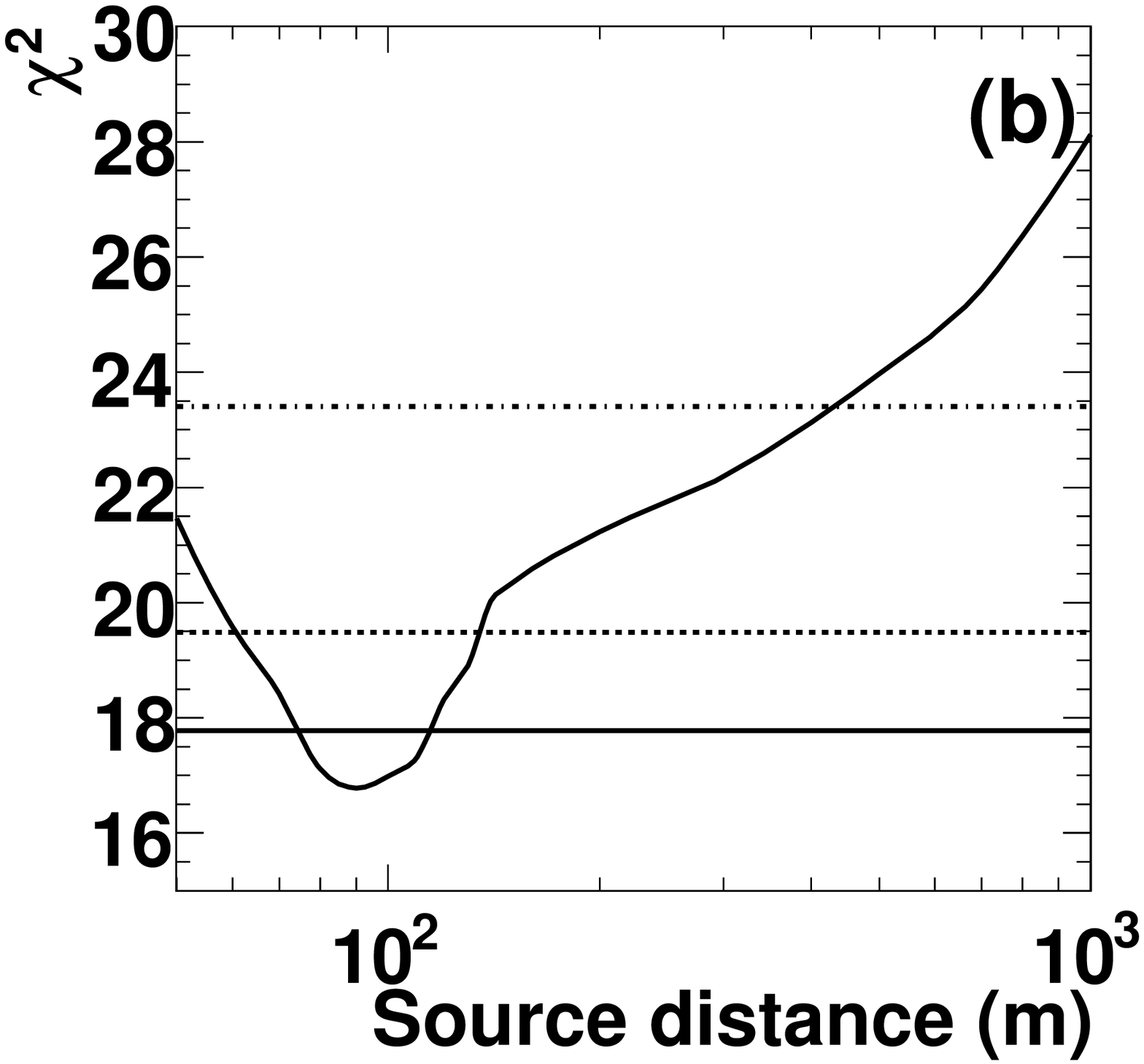}
\caption{%
(a) The background-subtracted photon energy spectra (black points) obtained from 
the NaI signals with anticoincidence. 
The horizontal and vertical axes represent the photon energy and the counts 
per energy bin, respectively. Error bars are statistical $1\sigma$.
Each curve shows prediction of an incident power-law model.
Air mass in $\mathrm{g\,cm^{-2}}$ corresponding to the assumed source distance is 
shown in parentheses.
(b) The $\chi^2$ value of the model fit to the data, plotted against  the 
assumed source distance. Horizontal lines from bottom to top represent 68\%, 90\%, and
99\% confidence levels.
}
\label{fig:spectra}
\end{figure}

Figure~\ref{fig:spectra}(a) shows the background-subtracted photon spectrum, 
summed over the burst period (90 sec).
Here, we subtracted a background spectrum averaged over 
15:30 $-$ 15:40 UT and 15:50 $-$ 16:00 UT, when the thunderstorm was ongoing.
This is to remove effects of Radon rain-outs, which gradually increased the $<3$ MeV
background up to twice those in quiescent periods.
The resultant background-subtracted spectrum 
exhibits a very hard continuum,  clearly extending to 10 MeV. 

We can estimate the initial photon energy spectrum at the source, 
from the background-subtracted spectrum. Theoretically, 
the number of non-thermal bremsstrahlung photons per unit energy
emitted by monochromatic electrons
is in inverse proportion to the photon energy, up to energies which are close to the 
electron energy~\cite{KM_1959}.  Also, a numerical simulation~\cite{Kotoku_2007}
has shown that a population of rather flat 
power-law distributed, or even monochromatic, relativistic electrons
colliding with a thick medium produces a power-law bremsstrahlung 
photon spectrum with a very hard photon index. 
Thus, we simply assume that the initial photon spectrum is a 
power-law of the form $\alpha  E^{-\beta}$, where $\alpha$ 
is a normalization factor  (photons $\mathrm{MeV^{-1}sr^{-1}}$), $\beta$ is the photon index, 
and $E$ is the photon energy in MeV. 
Then, to derive the photon energy spectra to be observed 
at the observatory, we simulated the photon propagation in the atmosphere with 
CORSIKA 6.500~\cite{CORSIKA} incorporating EGS4~\cite{EGS4} to correctly treat
electromagnetic interactions in relatively low energies relevant to the present work.
The photons were assumed to be injected vertically into the atmosphere, and
to propagate while suffering Compton scattering and other processes.
Finally,  the simulated arrival photon spectra, after convolving with the detector response, 
were fitted to the observed photon spectrum in search for
the best-fit values of $\alpha$ and $\beta$ for various source distances assumed.
The predicted spectra for representative source distances up to 1000 m 
are shown in Fig.~\ref{fig:spectra}(a); larger distances give poorer fits  (e.g. $\chi^2=42.2$ for 3000 m).
From Fig.~\ref{fig:spectra}(b),
the source distance, $d=90$ m, gives the minimum $\chi^{2}/d.o.f. = 16.8/16$, with 
$\alpha = (2.4\pm 0.4)\times 10^8\,\, \mathrm{MeV^{-1}sr^{-1}}$ and $\beta = 1.15\pm0.09$
(quoted errors are 90\% confidence values).
Figure \ref{fig:spectra}(b) also gives a constraint as $d = 60-130$ m at 90\% confidence level.

\section{Discussion\label{sec:dis}}
The present $\gamma$-ray burst is not due to Radon or its daughters,
because the $\gamma$-rays related to Radon have energies of 0.1 $-$ 2 MeV, 
and last for a much longer time~\cite{SUSZ_1996}.  
Compared with some prolonged intracloud x-ray emissions~\cite{MP_1985,Eack_1996},
this burst strongly suggests that electrons were accelerated beyond 10 MeV in thunderclouds.
Furthermore, the most outstanding difference of the present burst from previous 
ones~\cite{MP_1985,Eack_1996,SUSZ_1996,EAS_2000,Chub_2000,Alex_2002,monju_2002,nr_muraki_2004,growth_2007}
is that charged particles, most likely electrons, were simultaneously observed.
This difference may be attributable to the short distance, $\sim 90$ m,
of the present event. Actually, 
20 (10) MeV electrons have a range of 110 (60) m 
at an altitude of 2770 m, which is in good agreement with the constraint on
$d$ derived above.

We can construct the following picture of the 
present event. 
As predicted by \citet{RREA_model_1992},
seed electrons with energies of $>100$ keV,  produced by e.g. cosmic rays,  
are electrostatically accelerated in thunderclouds to relativistic energies,  and are
multiplied therein. We may expect the electron spectrum to reach $\sim$ 20 MeV,
because a numerical study~\cite{Roussel_2008} predicts that the runaway 
electrons have an average kinetic energy of $\sim$ 7 MeV with
a spread of $\sim$ 12 MeV. The 20 MeV electrons 
can effectively emit bremsstrahlung $\gamma$ rays extending to 10 MeV,
and escape from the thunderclouds to propagate a distance of $\sim$ 90 m, 
with their energies reducing down to
a few MeV. 
Because of a Compton optical depth of 0.51 (0.12) at 1 (10) MeV for $d=90$ m, 
the emitted $\gamma$ rays are sometimes scattered by large angles with significant energy
loss, producing secondary electrons.

If the primary electrons were in a perfect parallel beam,
relativistic effects would make
the emitted $\gamma$ rays, especially high-energy ones,
mostly beamed into a forward narrow cone with a half-opening angle of $\Gamma^{-1}\sim 1.5^\circ$,
where $\Gamma = 40$ is the Lorentz factor of 20 MeV electrons.
In practice, the cone will be much broader, due to multiple scatterings
of the emitting electrons, and to the mild Compton scatterings of the emitted photons.
The $\gamma$-ray spectrum is expected to be hardest along the cone axis.
Since the observed spectrum clearly extends to 10 MeV, it is thought that  
a rather harder part of the $\gamma$-ray emissions was detected.
In particular, the derived photon index, $1.15\pm0.09$, is close to the theoretically hardest 
limit (1.0). Thus, we infer that our detectors viewed the cone nearly along its axis.

We may estimate how the Compton-scattered electrons
and the escaping primary ones contribute to the excess counts detected
by the plastic scintillator, $N_\mathrm{ob}= 658 \pm 104$.
According to a Monte Carlo simulation, the survival probability
of $>1$ MeV (at the detector) secondary electrons, produced by 3, 5, and 10 MeV
$\gamma$ rays propagating from a source at $d = 90$ m, 
are 0.01, 0.02, and 0.04, respectively. 
Thus, employing the photon spectrum at $d=90$ m, 
$f_{90}(E)=2.4\times10^8E^{-1.15}$,
the expected count of the $> 1$ MeV secondary electrons in the plastic 
scintillator becomes at most $n_\mathrm{se}\sim 110$ even if all the scattered electrons
with arriving energies of $>1$ MeV are collected: 
we neglected lower-energy ones, since ionization losses in the 1 mm thick 
aluminum window prevents them from depositing $> 500$ keV energies in the plastic scintillator.
Hence,  we presume that the escaping primary electrons contribute to $N_\mathrm{ob}$ by
$N_\mathrm{d } = N_\mathrm{ob} - n_\mathrm{se} = 548\pm104$.

From this $N_\mathrm{d}$, we can evaluate the total number 
of 20 MeV primary electrons for $d=90$ m as 
\begin{equation}
N_\mathrm{20}=N_\mathrm{d}d^2/(S\epsilon)\sim 3.1\times10^8, 
\end{equation}
where $S = 1800\,\, \mathrm{cm^2}$ represents the area of the plastic scintillator, and 
$\epsilon \sim 0.08$ denotes the attenuation factor of $>1$ MeV electrons 
in the atmosphere and the aluminum window. Using this $N_\mathrm{20}$ and the initial photon 
spectrum $f_{90}(E)$, the spatial vertical length of the acceleration region, needed 
for the accelerated electrons to produce the observed bremsstrahlung $\gamma$ rays, 
can be estimated with an assumed beam half-opening angle $\theta_{h}$ as
\begin{equation}
H  \sim 2\pi N^{-1}_\mathrm{20}\int_{0}^{\theta_\mathrm{h}}\int _{1} ^{12} f_{90}(E)/\eta(E,\theta)\sin{\theta}dEd\theta,
\end{equation}
where $\eta(E,\theta)$ is the probability per 1 $\mathrm{g\,cm^{-2}}$ with which the 20 MeV electrons
emit bremsstrahlung photons with an energy $E$ and an angle $\theta$ with respect to the cone axis~\cite{KM_1959}. 
At $\theta_\mathrm{h}=15^\circ$, which is ten times larger than the perfect parallel 
beam case ($1.5^\circ$), the above $H$ becomes 200 m, and in turn
gives the electrostatic potential difference in the acceleration region 
as $U=200\,\, \mathrm{kVm^{-1}}\times 200\,\, \mathrm{m}=40 \,\, \mathrm{MV}$, 
where $200\,\, \mathrm{kVm^{-1}}$ is the critical electric field for seed electrons to cause the 
runaway electron avalanches given by
$280 P\,\, \mathrm{kVm^{-1}}$~\cite{Dwyer_MC2007} with the atmospheric pressure $P=0.72$ atm.
This $U$ is generally sufficient to accelerate high-energy seed electrons to relativistic energies or
$\sim$ 20 MeV if an electric field is higher than the critical value by $30-40$\%.
As discussed so far, this observation suggests that 
long-duration emissions of $\gamma$ rays and electrons are due to 
relativistic runaway electrons, although other possibilities may not necessarily be excluded.

\begin{acknowledgments}
We deeply thank the staff of Norikura cosmic-ray observatory of Institute for
Cosmic Ray Research, the University of Tokyo, for support of our experiment.  
We are grateful to Y. Ikegami and S. Shimoda 
for production of part of our system. 
The authors thank G. Poshak for carefully reading the manuscript.
The work is supported in part by 
Grant-in-Aid for Young Scientists (B), No. 19740167, 
the Special Podoctoral Research Project for Basic Science in RIKEN, 
and the Special Research Project for Basic Science in RIKEN (``Investigation of Spontaneously Evolving Systems"). 

\end{acknowledgments}
\newpage 

\clearpage
%
%
\clearpage
%
%
\clearpage
%
%

\end{document}